\documentclass[12pt]{iopart}
\usepackage{graphics}
\usepackage{bm}
\usepackage{epsfig}
\usepackage{citesort}
\usepackage{color}

\DeclareMathAlphabet{\mathpzc}{OT1}{pzc}{m}{it}

\begin{document}
\begin{flushright}
{\large \tt
 TTK-12-14}
\end{flushright}
\title{Origin of $\Delta N_{\rm eff}$ as a Result of an Interaction between Dark Radiation and Dark Matter}
\author{Ole Eggers Bjaelde}
\address{Department of Physics and Astronomy,
 University of Aarhus, 8000 Aarhus C, Denmark}
 \ead{\mailto{oeb@phys.au.dk}}
\author{Subinoy Das}
\address{Institut f\"ur Theoretische Teilchenphysik und Kosmologie \\ RWTH Aachen,
D-52056 Aachen, Germany, Department of Physics and Astronomy, University of British Columbia, 6224 Agricultural Road, Vancouver, BC V6T 1Z1, Canada}
\ead{\mailto{subinoy@physik.rwth-aachen.de}}
\author{Adam Moss}
\address{Department of Physics and Astronomy, University of Nottingham, Nottingham, NG7 2RD, UK}
\ead{\mailto{Adam.Moss@nottingham.ac.uk}}
\date{\today}

\begin{abstract}

Results from the Wilkinson Microwave Anisotropy Probe (WMAP), Atacama Cosmology Telescope (ACT) and recently from the South Pole Telescope (SPT) have indicated the possible existence of an extra radiation component in addition to the well known three neutrino species predicted by the Standard Model of particle physics. In this paper, we explore the possibility of the apparent extra \textit{dark} radiation being linked directly to the physics of cold dark matter (CDM). In particular, we consider a generic scenario where dark radiation, as a result of an interaction, is produced directly by a fraction of the dark matter density effectively decaying into dark radiation. At an early epoch when the dark matter density is negligible, as an obvious consequence, the density of dark radiation is also very small. As the Universe approaches matter radiation equality, the dark matter density starts to dominate thereby increasing the content of dark radiation and changing the expansion rate of the Universe.  As this increase in dark radiation content happens naturally after Big Bang Nucleosynthesis (BBN), it can relax the possible tension with lower values of radiation degrees of freedom measured from light element abundances compared to that of the CMB. We numerically confront this scenario with WMAP+ACT and WMAP+SPT data and derive an upper limit on the allowed fraction of dark matter decaying into dark radiation.

\end{abstract}

\maketitle
%%%%%%%%%%%%%%%%%%%%%%%%%%%%%%%%%%%%%%%%%%%%%%%%%%%%%%%%%
\section{Introduction}%%%%%%%%%%%%%%%%%%%%%%%%%%%%%%%%%%%
%%%%%%%%%%%%%%%%%%%%%%%%%%%%%%%%%%%%%%%%%%%%%%%%%%%%%%%%%

Measurements of fluctuations in the Cosmic Microwave Background (CMB) power spectra~\cite{Komatsu:2010fb} and light element abundances from Big Bang Nucleosynthesis (BBN)~\cite{Steigman:2007xt} have been corner stones in the era of precision cosmology. Though standard $\Lambda$CDM cosmology is a very good fit to present data from these measurements, there are also tantalising hints of physics beyond standard $\Lambda$CDM. One such example is the possible presence of an extra \textit{dark} radiation component during the epoch of decoupling. Recent analyses of CMB data point towards the possible existence of one or more radiation component(s)~\cite{Hamann:2007pi,Archidiacono:2011gq,Calabrese:2011hg,Hou:2011ec,Archidiacono:2012gv} with no standard electromagnetic and electroweak interactions other than those predicted by the standard model of particle physics. This extra radiation needs to be \textit{dark} in the sense that the presence of an extra photon like component would not only spoil the success of BBN, but also generate a chemical potential for the photon -- something which is constrained by CMB observations. The indication of excess radiation arises mainly through the precise observation of less power in the smaller scales of CMB anisotropy spectra.  It has also been confirmed that these hints for extra radiation are indeed `real', insofar as not being a statistical ambiguity from the choice of confidence interval~\cite{Hamann:2011hu}. Recently, the evidence was bolstered through the possible indication of ultra light sterile neutrino states in the neutrino oscillation experiments
\cite{Kopp:2011qd,Akhmedov:2010vy,Agarwalla:2010zu,Giunti:2011gz,Hamann:2010bk} and also through the reactor neutrino anomaly
\cite{Mention:2011rk}.

The nature of the dark radiation component is a topic of much debate. Current data allow the dark radiation component to be comprised by both sterile neutrinos as well as active neutrinos with a nonthermal distribution (see e.g. \cite{Cuoco:2005qr}). If dark radiation is comprised by massless sterile neutrinos, we expect them to behave as relativistic particles with effective sound speed $c_{\rm eff}^2$ and viscosity parameter $c_{\rm vis}^2$ satisfying $c_{\rm eff}^2=c_{\rm vis}^2=1/3$\footnote{Check definitions in \cite{Hu:1998kj}.}. Possible deviations from these values could indicate nonstandard interactions in the neutrino sector \cite{Beacom:2004yd,Hannestad:2004qu,Cuoco:2005qr,Basboll:2008fx,Basboll:2009qz}. Luckily, measurements of CMB anisotropies can help in constraining these parameters \cite{Hu:1998tk} and most analyses are consistent with the $c_{\rm eff}^2=c_{\rm vis}^2=1/3$ (see e.g. \cite{Trotta:2004ty,DeBernardis:2008ys,Archidiacono:2012gv}) - although \cite{Smith:2011es} reported on finding $c_{\rm eff}^2<1/3$. The bottom line is that cosmological data is sensitive to the details of dark radiation and can help in predicting its nature.

The helium abundance $Y_P$  is very sensitive to the expansion rate of the Universe (and hence the amount of radiation present) at the time when $ T \sim {\rm MeV}$. However, the evidence for extra radiation from BBN data is somewhat ambiguous. In some analyses, it is reported that one can accommodate one extra dark radiation component~\cite{Izotov:2010ca}, while in other works it is concluded that there is no need for extra radiation during BBN~\cite{Simha:2008zj,Aver:2010wq}. The former of these studies, for example, found $N_{\rm eff}^{\rm BBN} = 2.4 \pm 0.4$~\cite{Simha:2008zj}. The main reason for the confusion is that $N_{\rm eff}$ is highly sensitive to how the helium abundance is treated in the analysis~\cite{Nollett:2011aa}.  In general, from many data analyses, it remains a possibility that the central value for the number of relativistic degrees of freedom allowed by CMB data is higher than that of BBN. Taking CMB data alone, the estimate is  $N_{\rm eff}^{\rm CMB} = 5.3 \pm 1.3$ from the Wilkinson Microwave Anisotropy Probe (WMAP) 7-year with Atacama Cosmology Telescope (ACT) data~\cite{Dunkley:2010ge}. Combined WMAP and South Pole Telescope (SPT) data give a slightly lower value of $N_{\rm eff}^{\rm CMB} = 3.85 \pm 0.62$~\cite{Keisler:2011aw}. The addition of baryon acoustic oscillations (BAO) data and the measurement of the Hubble parameter $H_0$ improves these constraints somewhat. It is found that $N_{\rm eff}^{\rm CMB} = 4.56 \pm 0.75$ for WMAP+ACT+BAO+$H_0$~\cite{Dunkley:2010ge} and  $N_{\rm eff}^{\rm CMB} = 3.86 \pm 0.42$ for WMAP+SPT+BAO+$H_0$~\cite{Keisler:2011aw}. Taken at face value, the latter two results suggest $\sim 2\sigma$ evidence for extra relativistic species. The Planck satellite will dramatically increase the precision of the inferred value of $\Delta N_{\rm eff} \simeq 0.26$~\cite{Hamann:2007sb} and should be able to find a mismatch (if there is one) between  $N_{\rm eff}^{\rm CMB}$ and $N_{\rm eff}^{\rm BBN}$ at the level of $4-5 \sigma$~\cite{Hamann:2007pi}.

To explain the apparent radiation excess, one can, of course, just add a weakly interacting neutrino-like fermion by hand. However, then the question remains of explaining the origin of such a particle. For a recent particle physics model explaining the radiation excess with three flavours of light right-handed neutrinos, though, see \cite{Anchordoqui:2011nh}. See also \cite{Sikivie:2009qn,Sikivie:2011,Lundgren:2010} for another explanation of excess radiation during decoupling through the Bose-Einstein condensation of a coherently oscillating dark matter axion. We would like to point out that though LSND~\cite{Aguilar:2001ty} and MiNiBooNE~\cite{AguilarArevalo:2010wv} indicate the existence of one or more eV scale sterile neutrinos~\cite{Abazajian:2012ys}, which are excellent candidates for the excess radiation hinted by small scale CMB data, it is very hard to reconcile two eV scale sterile neutrinos as dark radiation with the large scale structure and other measurements \cite{Hamann:2011ge} unless sterile neutrinos have other interactions~\cite{Fan:2012ca,Barger:2003rt,Mangano:2010ei,Fardon:2003eh,Antusch:2008hj}. So, it is highly possible that the dark radiation may be a result of dark sector physics. For instance, if dark matter decays into dark radiation,  that can explain the dark radiation excess and its effect could be found in the structure formation of the Universe. Note that some hidden sector models \cite{Feng:2011uf,Blennow:2012de,Ichikawa:2007jv} motivated by other issues of particle physics and cosmology can also provide extra $\Delta N_{\rm eff}$. \\

If it is indeed the case that there is a change (increase) in the number of radiation degrees of freedom between the epoch of BBN and CMB, that will be an extremely interesting and surprising result. From a theoretical point of view, some new physics has to set in at a low energy scale $(T \sim {\rm eV})$. Recently, there have been a few interesting works in this line of thought \cite{Fischler:2010xz,Hasenkamp:2011em}, where $ \Delta N^{\rm BBN}_{\rm eff} \neq \Delta N^{\rm CMB}_{\rm eff}$. From a particle physics view point, this indicates that a particle (beyond the frame work of the standard model) has to decay \cite{hep-ph/0703034} into an extra dark radiation component in between the epoch of BBN and photon decoupling.

In this paper we propose a very simple mechanism where one naturally generates an extra radiation component when the Universe approaches the era of matter radiation equality and decoupling. The basic idea is to allow dark matter to interact with and decay into dark radiation: As the Universe approaches matter radiation equality (MRE), the density of dark matter starts to dominate the universal energy budget. As a result of the interaction, the densities of dark matter and dark radiation are proportional, hence the density of dark radiation increases as we approach MRE. In this scenario, one would naturally see an increase in the dark radiation component after BBN but before decoupling. One extra advantage of this scenario is the fact that the decay naturally reduces the amount of dark matter in galaxies and clusters. As a consequence it may help to alleviate \cite{Bell:2010fk,Abdelqader:2008wa} the well known small scale structure issues in $\Lambda$CDM cosmology - the problems with cuspy cores and overproduction of satellite galaxies in numerical simulations of structure formation \cite{Moore:1999nt,Klypin:1999uc,Navarro:1996gj}. We leave the details of this as possible future work.

In this paper we solve for the dark radiation density as a function of redshift numerically and show that we can obtain $\Delta N_{\rm eff}\rightarrow1$ as the Universe approaches the epoch of photon decoupling. We confront this generic scenario with the present cosmological data. We take a model-independent approach, where we use WMAP with either ACT or SPT data\footnote{We do not use ACT and SPT data simultaneously. This is because the observational fields slightly overlap and hence require a more detailed analysis of the combined noise properties. We thank Mark Halpern for pointing this out.} to constrain the fraction of the dark matter density which is allowed to be converted into dark radiation.  We show that one can easily find a viable region in parameter space where $\Delta N_{\rm eff}^{\rm CMB}$ can be greater than $\Delta N_{\rm eff}^{\rm BBN}$ by of order unity.

The plan of the paper is as follows:
In section~\ref{interaction}, we quantify the production of dark radiation and solve for the background solution. We present an analytical expression of $\Delta N_{\rm eff}$ and plot its dependence on scale factor and the coupling between dark matter and dark radiation. In section~\ref{perturbations}, we derive the cosmological perturbation equations for our scenario and, in section~\ref{cosmomc}, we discuss our main numerical results from a COSMOMC analysis using various datasets. We demonstrate that observations are consistent with $\Delta N_{\rm eff}\sim1$ around decoupling and $\Delta N_{\rm eff}\sim0$ around BBN. In section~\ref{origin}, we present a specific model in which the dark radiation production can be realised and constrain the model parameters. Finally we conclude in section~\ref{conclusion}.

%%%%%%%%%%%%%%%%%%%%%%%%%%%%%%%%%%%%%%%%%%%%%%%%%%%%%%%%%
\section{Interaction between dark radiation and dark matter} \label{interaction}
%%%%%%%%%%%%%%%%%%%%%%%%%%%%%%%%%%%%%%%%%%%%%%%%%%%%%%%%%

\subsection{Background evolution}

If dark radiation belongs to the dark sector along with dark matter, an interaction between the two could be possible. A general coupling (at the background level) can be described by the energy balance equations
\begin{eqnarray} \label{eq:contnu}
 \dot{{\rho}}_{\rm DM}&+&3H{\rho}_{\rm DM}=-Q\,,  \nonumber \\
  \dot{{\rho}}_{\rm dark}&+&3H \left( 1 + w_{\rm dark} \right) {\rho}_{\rm dark}= Q\,,
\label{eq:contDM}
\end{eqnarray}
%and
%\begin{equation}
%\centering
 %\label{eq:contnu}
%\end{equation}
where $\rho_{\rm DM}$ and $\rho_{\rm dark}$  are the dark matter/radiation energy densities and $H=\dot{a}/a$ is the Hubble rate, where $a$ is the scale factor and an overdot denotes the derivative with respect to conformal time $\tau$.  For our case of dark matter being converted into dark radiation then $w_{\rm dark}= P_{\rm dark}/\rho_{\rm dark} = 1/3$. The rate of energy transfer is given by $Q$ --  a positive $Q$ denotes the direction of energy transfer from dark matter to dark radiation. A non-zero $Q$ means that dark matter no longer redshifts exactly as $1/a^3$ and also that dark radiation does not redshift as $1/a^4$. It is important to remind ourselves that we require a covariant form for the energy momentum transfer $Q$.

Several papers over the recent years have studied different forms of the energy transfer rate $Q$ in the context of interacting dark matter-dark energy \cite{Zimdahl:2001ar,Wang:2005jx,Olivares:2006jr,Koivisto:2005nr,Ziaeepour:2003qs,Setare:2007we,Bjaelde:2007ki,Bjaelde:2008yd,Das:2005yj,Malik:2002jb,Valiviita:2008iv}. We  adopt the covariant form of energy momentum transfer 4-vector introduced in \cite{Valiviita:2008iv}
\begin{equation}
Q_{\rm DM}= \Gamma \rho_{\rm DM} \,,
\label{eq:covQ}
\end{equation}
where the form of interaction rate $\Gamma$ depends on the details of the particle physics of the decay process. Many forms of $\Gamma$  has been studied in literature, we adopt a simple case where $\Gamma= \alpha H $, where $\alpha$ is a constant and $H$ is the Hubble rate. As discussed in \cite{Valiviita:2008iv}, an implicit assumption behind this form of $\Gamma$ is that the interaction rate varies with time but not with space, which explains the presence of $H$ in the place of the interaction rate in Eq.~\ref{eq:covQ}. This form of $\Gamma$ can arise from different models of dark matter decay. In  section~\ref{origin}, we demonstrate a model of dark matter decaying into dark radiation and we show that  the above mentioned form of $\Gamma$ can be easily realised in nature.

For this form of the coupling it is easy to solve the background energy density equations~\cite{Valiviita:2008iv}
\begin{eqnarray}
\label{eq:rhodm}
\rho_{\rm DM} &=& \rho_{\rm DM,0} a^{-(3+\alpha)}\,,\nonumber\\
\rho_{\rm dark} &=& \rho_{\rm dark,0}a^{-3(1+w_{\rm dark})}+\left( \frac{\alpha}{\alpha-3w_{\rm dark}} \right) \rho_{\rm DM,0} a^{-3} (a^{-3w_{\rm dark}}-a^{-\alpha})\,.
\end{eqnarray}
With $w_{\rm dark}= 1/3 $ the equation for $\rho_{\rm dark}$ can be collected into two terms
\begin{equation}
 \rho_{\rm dark} = \beta a^{-4} + \left( \frac{\alpha}{1-\alpha} \right) \rho_{\rm DM,0} a^{-(3+\alpha)},
\label{eq:rhodark}
\end{equation}
where $\beta$ is a constant. The first part behaves like a standard radiation density and the second part behaves like a fluid with an equation of state $\alpha/3$. In the case of a weak coupling between dark matter and dark radiation, we require that $\alpha$ is small, which in turn leads to $\beta\sim0$. In the following we only keep the second term in Eq.~\ref{eq:rhodark}. This is further justified by the fact that the fraction of $\rho_{\rm dark}$, which redshifts like $1/a^4$, will be subdominant to the fraction which redshifts like dark matter due to the expansion of the Universe.

With this assumption we obtain
\begin{equation}
\frac{\rho_{\rm dark}}{ \rho_{\rm DM}} \to \frac{\alpha}{3w_{\rm dark}-\alpha} = \frac{\alpha}{1-\alpha}\,.
\label{eq:assumption}
\end{equation}
As we will see later from our numerical analysis, $ \alpha \ll 1$, so the ratio reduces to ${\rho_{\rm dark}}/{\rho_{\rm DM}}= \alpha  $. In fact, in section 6 we study a model of dark matter decay for a specific interaction and decay mechanism where our assumption is realised. This means that the pure radiation-like component (the term with coefficient $\beta$) is indeed absent.

\subsection{Calculation of $\Delta N_{\rm eff}$ }

In the standard cosmological scenario, it is a standard practice to define $\Delta N_{\rm eff}$ by
\begin{equation}
  \rho_{\rm rad} = \left[ 1 + \frac{7}{8}  N_{\rm eff} \left( \frac{T_{\nu}}{T_{\gamma}} \right)^4 \right]  \rho_{\gamma}\,,
  \label{rhorad}
\end{equation}
where the radiation  density $\rho_{\rm rad}$ is given as a sum of the energy density in photons $\rho_{\gamma}=(\pi^2/15) T_{\gamma}^4$ and standard model neutrinos. In the standard model this predicts $N_{\rm eff}^{\rm SM}=3.046$ with $T_{\nu}/T_{\gamma} = (4/11)^{1/3}$. Any departure from the standard scenario is parameterized as $N_{\rm eff} = N_{\rm eff}^{\rm SM} + \Delta N_{\rm eff}$.

\begin{figure}[!htb]
 \begin{center}
 \includegraphics[width=6.0 in]{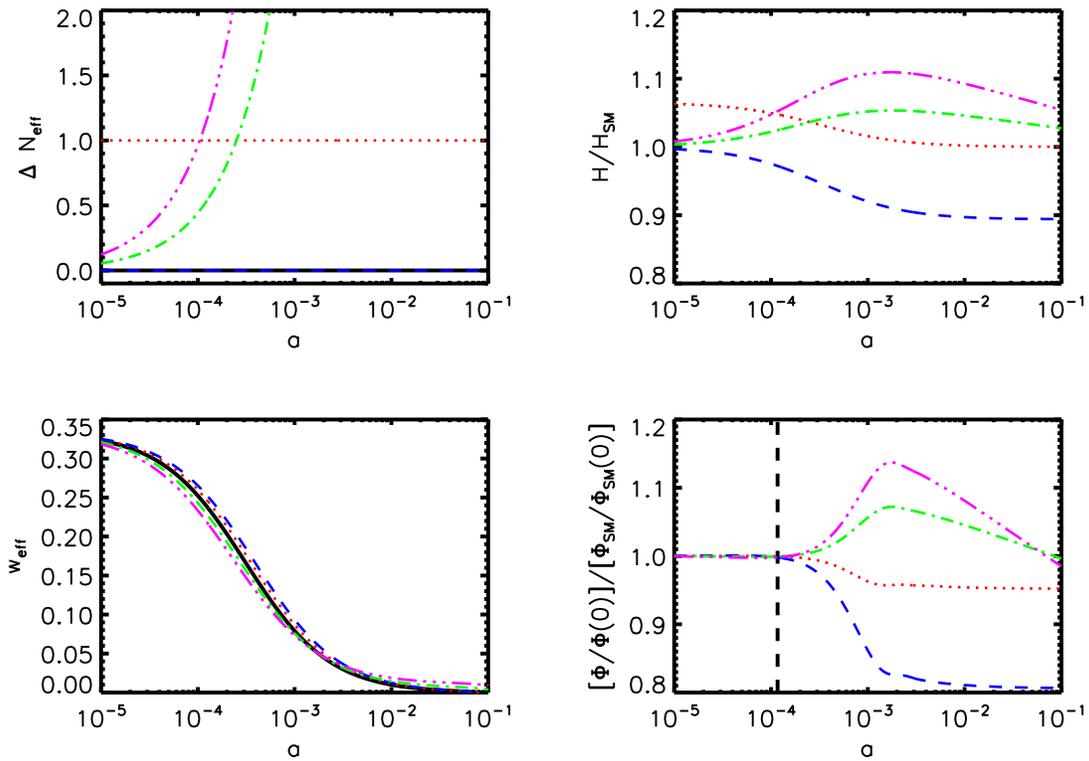}
 \caption{For each panel we show the best-fit vanilla 6-parameter model from WMAP+SPT  (black), then models with the {\em same} parameters but  one extra relativistic species (dotted-red), a lower dark matter density of $\Omega_{\rm DM} = 0.085$ (dashed-blue, as opposed to  $\Omega_{\rm DM} = 0.112$) and a decaying dark matter model with $\alpha = 0.02$ (dot-dash green) and 0.04 (dot-dot-dash magenta). (Top-left) $\Delta N_{\rm eff}$ as a function of scale factor. (Top-right) Hubble rate compared to the standard model. (Bottom-left) Effective (total) equation of state. (Bottom-right) Ratio of the gravitational potential $\Phi$ for a Fourier mode with $k=0.02\, {\rm Mpc}^{-1}$ compared to the standard model. Horizon entry for this mode is indicated by the dashed vertical line.}
 \label{fig:plots}
 \end{center}
\end{figure}

One implicit assumption in the above definition is that the dark matter dilutes as $1/a^3$. But in our case neither dark matter nor dark radiation dilutes in the standard way. This means we cannot simply use the definition above -- a more appropriate model independent method is needed to compare the expansion rate $H$ to that of the standard model $H_{\rm SM}$ and attribute the difference to $ \Delta N_{\rm eff}$. In our model
\begin{eqnarray}
3 H^2 M_{Pl}^2 &=& \rho_{\rm DM,0} / a^{3 + \alpha} + \frac{\alpha}{1-\alpha} \rho_{\rm DM,0}/a^{3 + \alpha} + \, \rho_{\rm rest} \\ \nonumber &=& \frac{1}{1-\alpha}  \rho_{\rm DM,0}/a^{3 + \alpha} + \, \rho_{\rm rest} \,,
\end{eqnarray}
where $\rho_{\rm rest}$ stands for the normal radiation and dark energy components. We then compare to the standard $H_{\rm SM}$ (with non-zero $\Delta N_{\rm eff}$) to obtain
\begin{equation}
 \frac{7}{8}  \Delta N_{\rm eff} \, \left (\frac{T_{\nu}}{T_{\gamma}} \right)^4  \, \frac{\rho_{\gamma 0}}{a^4} =   \frac{1}{1-\alpha}  \rho_{\rm DM,0}/a^{3 + \alpha} - \rho_{\rm DM,0}/a^{3}\,.
\end{equation}
Using the above definition we find that $\Delta N_{\rm eff}$ depends on the decay constant $\alpha$ as well as scale factor $a$. We show $\Delta N_{\rm eff}$ as function of $a$ in the top-left panel of Fig.~\ref{fig:plots} for  $\alpha=0.02$ and 0.04. At the time of decoupling this produces a $\Delta N_{\rm eff}$ of order unity.

As already discussed, the contribution to $\Delta N_{\rm eff}$ arises mainly due to the faster expansion rate at the time of decoupling and MRE in our model. It is interesting to note that at an early epoch deep in the radiation-dominated era, when the dark matter density is negligible, the deviation from the standard expansion rate is close to zero, as expected. At late times -- after $ z \sim 100$ -- the expansion rate again approaches the standard model expansion rate, as demonstrated in Fig.~\ref{fig:plots}. This is because $ a \rightarrow 1$ and $\alpha$ is still small -- in other words the interacting dark matter starts to behave in the same way as in standard $\Lambda$CDM.

%%%%%%%%%%%%%%%%%%%%%%%%%%%%%%%%%%%%%%%%%%%%%%%%%%%%%%%%%
\section{Perturbations of dark radiation}    \label{perturbations}
%%%%%%%%%%%%%%%%%%%%%%%%%%%%%%%%%%%%%%%%%%%%%%%%%%%%%%%%%

To determine the perturbations to dark radiation and dark matter we need to consider the Boltzmann equation for its distribution. For simplicity we will consider a massive dark matter particle decaying into a pair of massless daughter particles. Following standard practice (e.g.~\cite{Ma:1995ey}) we expand the distribution function for each species $j$ in terms of a zero-order component $f^0_j$ and a perturbation $\Psi_j$
\begin{equation}
 f_j(x^i,q_j,n_i,\tau) = f_j^0(q_j,\tau)[1+\Psi_j(x^i,q_j,n_i,\tau)]\,,
 \end{equation}
which depends on position $x^i$, magnitude of momentum $q_j$, direction $n_i$ and conformal time $\tau$. The phase space
of each species obeys the Boltzmann equation
\begin{equation}
 \frac{Df_j}{d\tau}=\frac{\partial f_j}{ \partial \tau}+\frac{\partial f_j}{\partial x^i}\frac{dx^i}{d\tau}
 +\frac{\partial f_j}{\partial q_j}\frac{dq_j}{d\tau}+\frac{\partial f_j}{\partial
 n_i}\frac{dn_i}{d\tau}=\left(\frac{df_j}{d\tau}\right)_C\,,
\end{equation}
where $\left(\frac{df_j}{d\tau}\right)_C$ is the collision term, which
depends on particle interactions.

At zeroth-order the Boltzmann equation for the dark matter distribution function can then be written as~\cite{Kawasaki:1992kg,Kaplinghat:1999xy}
\begin{equation}
  \dot{f}_{\rm DM}^0=-\alpha H f_{\rm DM}^0\,,
  \label{eq:collision}
\end{equation}
under the assumption that Eq.~\ref{eq:assumption} is fulfilled. Upon multiplying by the proper energy $\epsilon_j = \sqrt{q_j^2 + a^2 m_j^2}$ and integrating over all momenta one obtains the same continuity equation for dark matter as in Eq.~\ref{eq:contDM}.

Working in the synchronous gauge, we can now work out the equations of motion for the perturbations to dark matter and its decay product. For the perturbations to dark matter we write out the Boltzmann equation and the perturbation to the energy density by following the machinery described in Ref.~\cite{Ma:1995ey}. In the end, the equations of motion for the dark matter perturbations reduce to the case of stable dark matter particles - the only difference being the different distribution function $f_{\rm DM}^0$ specified by Eq.~\ref{eq:collision}. This result was also obtained in Refs.~\cite{Kaplinghat:1999xy,Wang:2010ma,Wang:2012ek}

Following Ref.~\cite{Ma:1995ey}, for the massless decay product we integrate out the $q$ dependence of the distribution function and expand the angular component in terms of Legendre polynomials,
\begin{equation}
\label{fsubl}
      F_j(\vec{k},\hat{n},\tau) \equiv {\int q_j^2 dq_j\,q_j f^0_j(q_j)\Psi
    \over \int q_j^2 dq_j\,q_j f^0_j(q_j)} \equiv \sum_{l=0}^\infty(-i)^l
    (2l+1)F_{j\,l}(\vec{k},\tau)P_l(\hat{k}\cdot\hat{n})\,,
\end{equation}
where $\mu=\hat{k}\cdot\hat{n}$ and $P_n(\mu)$ are the Legendre
polynomials of order $n$.

The Boltzmann equation can then be worked out to give
\begin{equation}  \label{frddot}
\dot{F}_{\rm dark} + i k \mu F_{\rm dark} = -\frac{2}{3} \dot{h} - \frac{4}{3} \left(\dot{h} + 6 \dot{\eta} \right) P_2 (\mu) +    H(1-\alpha)  \left( \delta_{\rm DM} - F_{\rm dark} \right)\,,
\end{equation}
where the expression for the density perturbation $\delta=\delta \rho/\rho$ in terms of phase-space integrals, and the definition of the synchronous gauge metric perturbations $h$ and $\eta$ can be found in  Ref.~\cite{Ma:1995ey}.

Eq.~\ref{frddot} can be translated into a hierarchy of perturbation equations of motion for
the dark radiation by inserting the expansion in Eq.~\ref{fsubl} and
collecting terms. The final results are
\begin{eqnarray}
\label{massless}
    \dot{\delta}_{\rm dark} &=& -{4\over 3}\theta_{\rm dark}
        -{2\over 3}\dot{h}-H(1-\alpha)(\delta_{\rm dark}-\delta_{\rm DM}) \,,\\
    \dot{\theta}_{\rm dark} &=& k^2 \left(\frac{1}{4}\delta_{\rm dark}
        - \sigma_{\rm dark} \right)-H(1-\alpha)\theta_{\rm dark}\,,\nonumber\\
    \dot{F}_{{\rm dark}\,2} &=& 2\dot\sigma_{\rm dark} = {8\over15}\theta_s
            - {3\over 5} k F_{{\rm dark}\,3} + {4\over15}\dot{h}
        + {8\over5} \dot{\eta}-H(1-\alpha)F_{{\rm dark}\,2} \,,\nonumber\\
    \dot{F}_{{\rm dark}\,l} &=& {k\over2l+1}\left[ l
        F_{{\rm dark}\,(l-1)} - (l+1)
        F_{{\rm dark}\,(l+1)} \right]-H(1-\alpha)F_{{\rm dark}\,l}\,, \quad l \geq 3 \,, \nonumber
\end{eqnarray}
where $\delta_{\rm dark} = F_{\rm dark \,0}$, $\theta_{\rm dark} = 3k/4 F_{\rm dark\,1}$ and $\sigma_{\rm dark} = F_{\rm dark\,2}/2$. This is an infinite hierarchy so we need to truncate the hierarchy at some $l_{\rm
max}$, for which we choose~\cite{Ma:1995ey}
\begin{equation}
\label{truncnu}
     F_{\nu\,(l_{\rm max}+1)}\approx{(2l_{\rm max}+1)\over k\tau}\,F_{\nu
       \,l_{\rm max}}-F_{\nu\,(l_{\rm max}-1)}\ .
\end{equation}
These equations are equivalent to those in Ref.~\cite{Kaplinghat:1999xy} for their choice of decay variables.

We implemented these equations in a modified version of  {\tt CAMB}~\cite{CAMB}. This amounts to: (1) Changing the scaling behaviour of dark matter to $a^{-(3+\alpha)}$; (2) Modifying the background evolution to include an additional component whose energy density scales like dark matter but with $w_{\rm dark} = 1/3$ ; (3) Implementing the hierarchy of perturbation equations, which are similar (with the exception of the final terms in Eq.~\ref{massless}) to the existing massless neutrino perturbation equations.

%%%%%%%%%%%%%%%%%%%%%%%%%%%%%%%%%%%%%%%%%%%%%%%%%%%%%%%%%
\section{Results} \label{cosmomc}
%%%%%%%%%%%%%%%%%%%%%%%%%%%%%%%%%%%%%%%%%%%%%%%%%%%%%%%%%

\begin{figure}[!htb]
\begin{center}
 \includegraphics[width=4.2 in]{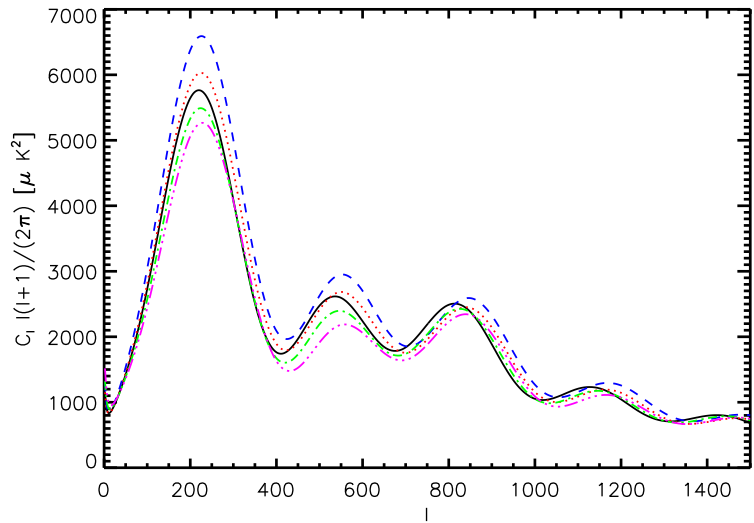}
 \caption{Temperature power spectrum for the models listed in Fig.~\ref{fig:plots}. }
 \label{fig:cl}
 \end{center}
\end{figure}

In Fig.~\ref{fig:cl} we show the temperature power spectrum each of the models listed in Fig.~\ref{fig:plots}. The general features can be understood qualitatively by the following

\begin{itemize}
\item For models with an extra relativistic species, lower CDM and decaying dark matter (DDM) the acoustic peaks are shifted to smaller scales. The angular scale of the peaks is set by the ratio of the sound horizon at decoupling to the angular diameter distance to decoupling. For a flat Universe this is approximately the  ratio of the conformal time at decoupling to that today, i.e. $\theta_{\rm A} \approx \tau_{\rm dec}/\tau_0$. For extra relativistic species the increased Hubble rate at early times decreases $\tau_{\rm dec}$, while $\tau_0$ remains similar. For lower CDM both $\tau_{\rm dec}$ and $\tau_0$ increase, with the relative increase in  $\tau_0$ compared to the standard model greater. For DDM both $\tau_{\rm dec}$ and $\tau_0$ decrease, with the relative decrease in  $\tau_{\rm dec}$ greater.
\item The first two peaks are noticeably enhanced in models with an extra relativistic species and lower CDM, but are suppressed in the DDM model. This arises from the driving effect (modes entering the horizon during the radiation era are enhanced due to the decay of gravitational potentials) and (for the first peak) the early Integrated Sachs-Wolfe (ISW) effect. The total (effective) equation of state of the Universe for each model is shown in Fig.~\ref{fig:plots}. For extra relativistic species and lower CDM, radiation domination ($w_{\rm eff} \approx 1/3$) is extended, while for the decaying dark matter model the Universe actually departs from radiation domination faster. This is somewhat opposite to what one might expect, since dark matter is decaying into dark radiation, but is a consequence of fixing $\Omega_{\rm DM}$ today to be the same as the standard model and it scaling as $a^{-(3+\alpha)}$. The result of this can be seen in Fig.~\ref{fig:plots}, by plotting the gravitational potential $\Phi$ for a mode with $k=0.02 \, {\rm Mpc}^{-1}$, which enters the horizon around $a \approx 10^{-4}$. Potential decay is suppressed in the DDM model by the time of decoupling.
\item The first peak is also affected by the early ISW -- here the potential can still decay after decoupling as the Universe is not completely matter dominated. For DDM the total equation of state is closer to the standard model at decoupling than in models with an extra relativistic species or lower CDM, so the early ISW contribution is smaller. There is, however, a late time ISW effect, resulting in more power on large scales. This is because the total equation of state $w_{\rm eff} \ne 0$, and also because the radiation decay product (whose energy density is $\alpha \rho_{\rm DM}$) contributes a source of pressure and anisotropic stress.
\item On small scales there is reduced power in the damping tail of the CMB for both extra relativistic species and DDM. This is due to the increased expansion rate prior to decoupling, resulting in higher diffusion damping. The opposite occurs for lower CDM due to the decreased expansion rate.
\end{itemize}

In order to confront the DDM  model with observations we perform parameter estimation using a modified version of the COSMOMC package~\cite{cosmomc}. For our analysis we use data from the 7-year WMAP release~\cite{Komatsu:2010fb}, the 148 GHz 2008 ACT data~\cite{Dunkley:2010ge} and the  150 GHz 2008/2009 SPT data~\cite{Keisler:2011aw}. Due to the small overlapping sky coverage between ACT and SPT we consider WMAP + ACT and WMAP + SPT independently.  We use software provided by each team to compute the likelihood of cosmological models.

\begin{figure}[!h]
\begin{center}
 \includegraphics[width=5.2 in]{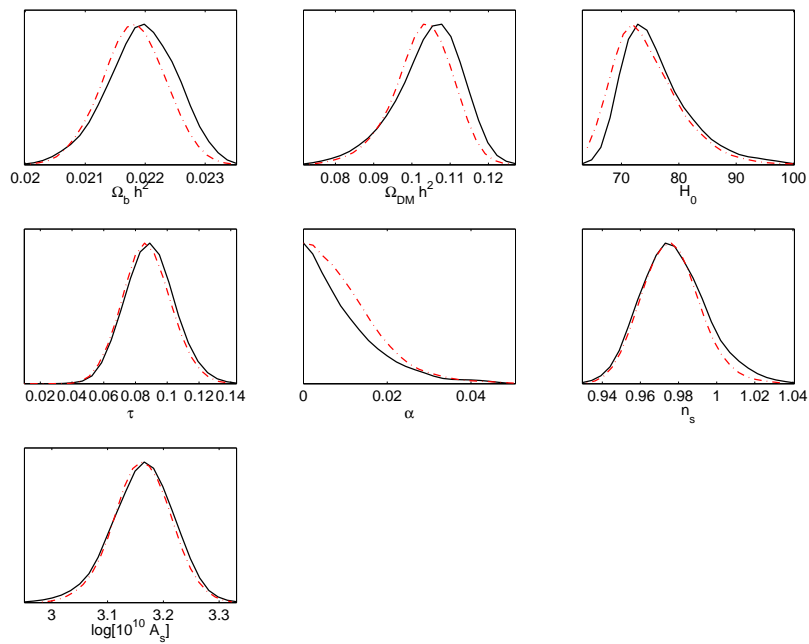}
 \caption{Marginalized parameter constraints for WMAP + ACT (solid black) and WMAP + SPT (dashed red).}
 \label{fig:1d}
 \end{center}
\end{figure}

To parameterize these models we fit for 7 parameters, imposing the flatness condition $\Omega_{\rm k}=0$\footnote{The effect of leaving $\Omega_{\rm k}$ as a free parameter has been investigated in e.g. \cite{Kristiansen:2011mp,Giusarma:2011zq}. Interestingly, in a model with two sterile neutrinos the cosmological constant seems to be ruled out at 95 \% confidence level. Furthermore, models with sterile neutrinos seem to prefer $w<-1$ for the dark energy equation of state.}: the baryon density $\Omega_{\rm b} h^2$, cold dark matter density $\Omega_{\rm DM} h^2$, Hubble parameter $H_0 = 100 \, h \, {\rm Mpc}^{-1} \, {\rm km} \, {\rm s}^{-1}$, optical depth to reionization $\tau$, and the amplitude $A_{\rm s}$ and spectral index $n_{\rm s}$ of initial fluctuations.  In addition we fit for the dark matter decay constant $\alpha$, imposing a prior that $\alpha \ge 0$. Since ACT and SPT observe much smaller scales than WMAP, marginalisation over foregrounds is also required, since these contribute to the small scale  temperature power spectrum.  We follow the same procedure as in the ACT and SPT analysis, marginalising over a combined thermal and kinetic Sunyaev-Zeldovich (SZ), and a clustered point source, template, together with a Poisson ($C_{\ell} = {\rm const}$) point source component.  The reader is referred to these references for more details on the templates used. This brings the total number of parameters fitted to 10 (7 cosmological and 3 secondary foreground parameters). We used the lensed theoretical CMB spectra from {\tt CAMB} in our fits, since lensing is favoured by WMAP + ACT/SPT at the level of several $\sigma$.

\begin{figure}[!t]
\begin{center}
 \includegraphics[width=5.2 in]{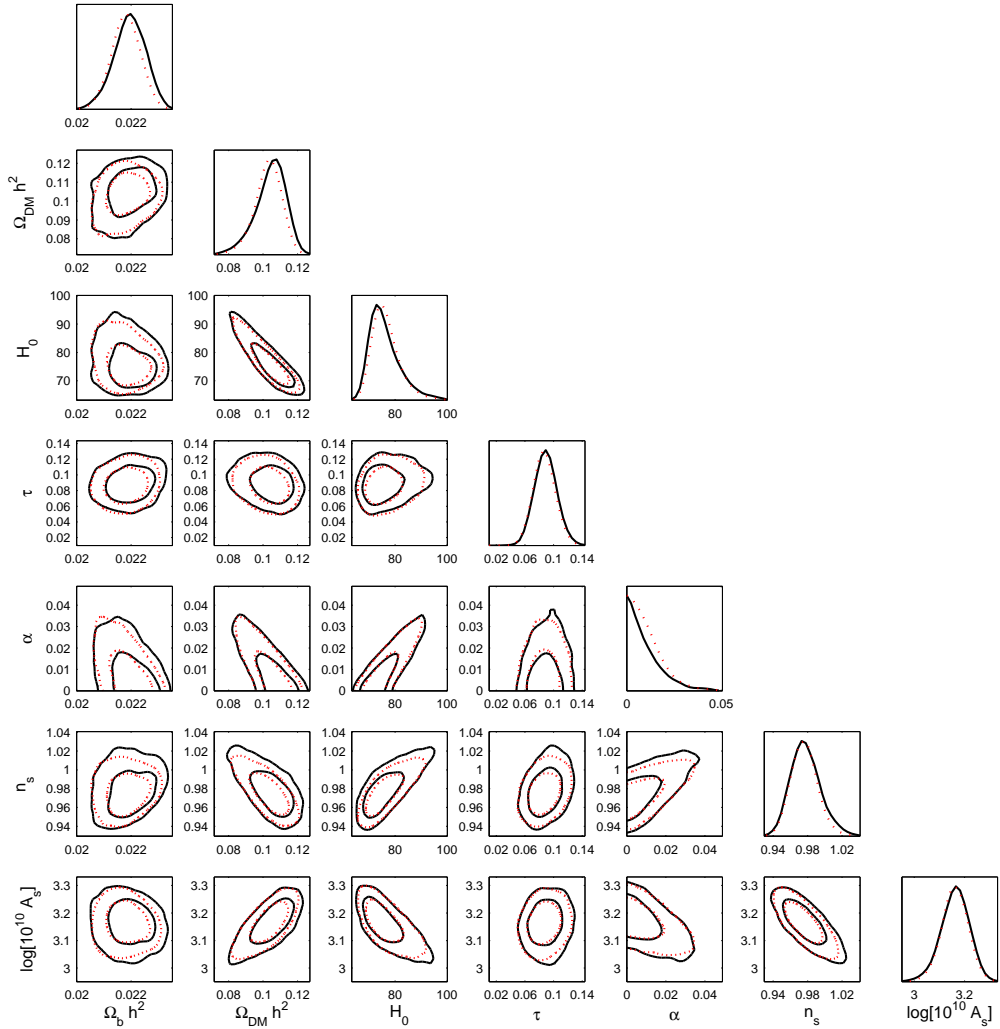}
 \caption{Marginalized constraints on the 7 fitted cosmological parameters for WMAP + ACT (solid black) and WMAP + SPT (dashed red). Likelihood contours show the 68\% and 95\% confidence levels.}
 \label{fig:contour}
 \end{center}
\end{figure}

The results of our analysis are shown in Fig.~\ref{fig:1d} and Fig.~\ref{fig:contour}, where we show marginalised 1-dimensional and 2-dimensional parameter constraints. There is no preference for a non-zero decay constant,  with a $2\sigma$ upper-limit of $\alpha< 0.027$ for WMAP + ACT and $\alpha< 0.028$ for WMAP + SPT. This likely stems from the different effect on the CMB spectrum for DDM than for extra relativistic species, in particular the suppression of the first peak instead of an enhancement. There is some degeneracy with other parameters, in particular the spectral index and Hubble parameter. Repeating our analysis for the standard model ($\alpha=0$) we find, for example, WMAP + SPT gives $n_{\rm s} = 0.964 \pm 0.011$ and $h=0.706 \pm 0.021$, while allowing $\alpha$ to vary then $n_{\rm s} = 0.976 \pm 0.015$ and $h=0.765 \pm 0.052$.

\section{Origin of interaction: A phenomenological model} \label{origin}

An empirical form of energy transfer has been introduced in most studies \cite{Zimdahl:2001ar,Wang:2005jx,Olivares:2006jr,Koivisto:2005nr,Ziaeepour:2003qs,Setare:2007we,Bjaelde:2007ki,Bjaelde:2008yd,Das:2005yj,Malik:2002jb,Valiviita:2008iv} where interactions between different cosmological sectors has been considered. Till now we also have chosen an empirical form $\Gamma = \alpha \, H$ for our model and used it to confront with data. For this form of coupling we have shown that the ratio of the energy densities of dark radiation and dark matter $\rho_{\rm dark} / \rho_{DM}= \alpha$ is practically constant in time for small $\alpha$. Hence, putting an upper bound on $\alpha$ basically gives us an estimate of how large a fraction of dark matter is allowed to decay into dark radiation, obeying all cosmological constraints.

Though the main goal of this paper is not a dark matter model, we will now discuss a specific phenomenological model where one can analytically derive the energy transfer equations between dark matter and dark radiation. We show that a time independent $\alpha$ can indeed emerge in a phenomenological model. Note, however, that our numerical results are not limited to this specific model. Any dark matter model with a coupling to dark radiation and where the fraction of dark radiation to dark matter does not change much in the course of a Hubble time will be subject to the constraints from section \ref{cosmomc}.\\

We consider a coherently oscillating scalar field which plays the role of CDM \cite{Das:2006ht,Bjaelde:2010vt,Kobayashi:2011hp}. This has a Yukawa type coupling to a nearly massless \textit{dark} fermion $\psi_d$\,\, (${\cal L} \supset \lambda \phi \psi_d \bar{\psi_d}$). This type of CDM can in principle decay parametrically into dark radiation and the situation is very similar to the fermionic preheating scenario in the context of inflation \cite{Greene:2000ew}. However, the energy scale which we are considering here is much lower compared to that of inflation. We refer our readers to Refs.~\cite{Das:2006ht,Bjaelde:2010vt,Kobayashi:2011hp} for details of the dark matter decay process in this scenario.
Here we present a brief review about the basic mechanism of \textit{dark} radiation production from CDM and finally we put constaints on the model parameters on the basis of the numerical results obtained in section \ref{cosmomc}.

By adopting the results of fermionic preheating \cite{Bjaelde:2010vt,Greene:2000ew} in an expanding background, the comoving number density of dark fermions can be found by solving the well known Mathieu equations
\begin{equation}
X_k^{''} + [\kappa^2 + (\tilde{m} + \sqrt{q} f)^2 -i \sqrt{q} f']
X_k =0,
\end{equation}
where the resonance parameter $q\equiv \lambda^2 \phi_0^2 / m_{\phi}^2$, $\phi_{0}f(t)$ is the background solution for the time
evolution of the oscillating scalar field, $\kappa \equiv
k / m_{\phi}$ is the dimensionless fermion mass, and $\tilde{m}
\equiv m_{\psi} / m_{\phi}$. These three
parameters completely determine the parametric production of
fermions. We consider the oscillation of the field with the usual
quadratic potential $V= \frac{1}{2} m^2 \phi^2 $, which is a good
first order approximation around the minima of any potential. The term $(\tilde{m} +
\sqrt{q} f)$ can be thought of as an effective mass of the fermion.
As the scalar field oscillates, the effective mass itself will
oscillate around zero and the parametric production of fermions is
enhanced when the effective mass crosses zero. It can be shown
numerically that $n_k(t)$ oscillates and due to Pauli blocking its
maximum value never crosses unity.\\

The expansion of the Universe has not been taken into account in the discussion above. As the mass of the oscillating scalar is very low, for a complete treatment, however, we have to consider the expansion of the Universe. When the expansion is taken into account, the resonance parameter
 $q \equiv \lambda^2 \phi (t)^2 / m_{\phi}^2 $ becomes time-dependent and the periodic modulation of the comoving number density does not hold any more. It has been shown \cite{Bjaelde:2010vt} (though in a different context of neutrino cosmology) that the parametric production of dark radiation happens as long as the time dependent resonance parameter satisfies $q(z) \gg 1$. In this regime, the produced dark radiation density takes a very simple form $\rho_{\rm dark} = 8 \pi \lambda^2 \rho_{\rm DM}$, which gives $\alpha$ in terms of model parameters $\alpha = 8 \pi \lambda^2$.
The only requirement for the above relation is $q \gg 1$, which translates into
\begin{equation}
2 \, \frac{\lambda^2 \, \rho_{\rm DM, 0}}{m_{\phi}^4} \, (1+z)^3 \gg 1,
\label{eq:lambdacon}
\end{equation}
which we assumes holds true from BBN till the present epoch denoted by the '0' superscript.

\begin{figure}[!htb]
 \begin{center}
 \includegraphics[width=3.2 in]{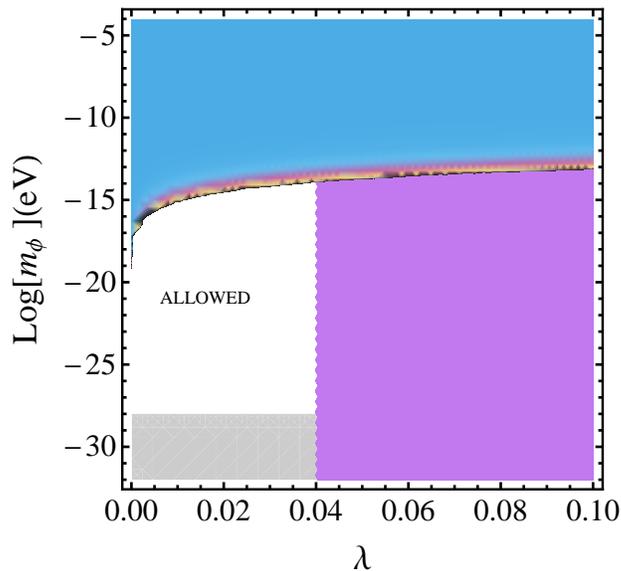}
 \caption{Allowed  region in the $( m_{\phi}, \lambda)$ plane for a coherently oscillating scalar dark matter decaying into dark radiation. The blue area is excluded from the requirement that the resonance parameter is not high enough to produce dark radiation through parametric resonance. The purple region is excluded from the upper limit on the fraction of dark matter decaying into dark radiation taken from our numerical result in section \ref{cosmomc}. The lower grey area is excluded from the requirement
 $m_{\phi} \gg H (T_{\rm osc}) $. }
 \label{fig5}
 \end{center}
\end{figure}

So far the dark matter mass $m_{\phi}$ has not been constrained in our analysis. This is because the dark matter mass does not enter explicitly into the perturbation analysis and thus our numerical results do not constrain the dark matter mass directly. One can, however, place an upper bound on the dark matter mass for this specific model of parametric decay. This is possible from the requirement  $m_{\phi} \geq H$ -- otherwise the Hubble friction would prevent the field from oscillating coherently and will not allow it to behave as CDM. From the constraints on CDM matter power spectra we know that dark matter has to be present in the Universe at least couple of e-foldings before MRE -- otherwise there would be too much suppression in the linear matter power-spectra on small scales \cite{Das:2006ht}. So to get an estimate we assume the scalar started to oscillate coherently when the temperature of the Universe was around  $T=T_{\rm osc} \simeq 100 \,$ eV. This choice keeps us out of the conflict with constraints from linear matter power spectra measurements from Lyman-$\alpha$ and SDSS data \cite{McDonald:2004eu,Viel:2005ha}. This, in turn, constrains the dark matter mass $ m_{\phi} \gg H(T_{\rm osc}) $.
In Fig.~\ref{fig5} we show the allowed region in the $(m_{\phi}, \lambda)$ plane which satisfy all of the above three constraints namely: a) To satisfy the parametric production of dark radiation; b) To be consistent with our numerical results (upper bound on $\alpha$) from section \ref{cosmomc}; c) To obey the condition for the coherent oscillation setting in before MRE.

%%%%%%%%%%%%%%%%%%%%%%%%%%%%%%%%%%%%%%%%%%%%%%%%%%%%%%%%%
\section{Discussion and conclusion} \label{conclusion}
%%%%%%%%%%%%%%%%%%%%%%%%%%%%%%%%%%%%%%%%%%%%%%%%%%%%%%%%\\	

The evidence for the existence of dark radiation at the CMB epoch is intriguing. If future experiments find a mismatch between the radiation content of the Universe at the epoch of BBN and decoupling, the production of dark radiation may be a late-time phenomenon in the cosmic history which took place some e-foldings after BBN. Future surveys like Planck will measure the effective number of radiation degrees of freedom with an accuracy of  $ \Delta N_{eff} = 0.026 $ \cite{Hamann:2007sb} and will also be able to probe if extra radiation has been produced after BBN at all, so that $ \Delta N^{\rm BBN}_{\rm eff} \neq \Delta N^{\rm CMB}_{\rm eff}$.

In this paper, we have shown that if dark radiation is produced from dark matter decays, the Universe naturally gets populated with an extra radiation component after BBN but before photon decoupling. The reason is that as the Universe cools the dark matter density increases and, as a result, so does the dark radiation produced from it. We have constrained the fraction of dark matter which is allowed to be converted into dark radiation using the WMAP7 + ACT and WMAP + SPT data. We find an upper bound on this fraction using a COSMOMC analysis and show that it is possible to get an increase in $N_{\rm eff}$ by of order unity as the Universe approaches the epoch of photon decoupling. However, the effect on the temperature power spectrum is somewhat different than adding in extra relativistic species by hand, most noticeably in the suppression of the first acoustic peak. For this reason,  if Planck confirms the mismatch between $\Delta N^{\rm CMB}_{\rm eff}$ and $\Delta N^{\rm BBN}_{\rm eff}$, it remains to be seen how well the decaying dark matter model fits data.

As a phenomenological example we have presented a model of dark matter decay and calculated the decay rate as a function of the coupling between CDM and dark radiation.\\

Dark matter decaying into dark radiation could also have important implications for cosmological structure formation.
Very recently the observations of high redshift massive galaxy clusters \cite{arXiv:1006.5639} has put $\Lambda$CDM cosmology under stringent constraints \cite{arXiv:1009.3884,arXiv:1006.1950,Mortonson:2010mj}. In fact, the presence of extra dark radiation during the CMB epoch may play a role in resolving these issues \cite{Bashinsky:2003tk}. Note in this context that in Fig.~\ref{fig:plots} we have showed that the decay of the gravitational potential during decoupling is suppressed in the decaying dark matter model. This may boost early structure formation as discussed above.
The detailed study of this effect in the context of our model is beyond the scope of this paper and we leave it for future work.\\

\ack

We thank Jim Zibin and Yvonne Wong for useful discussions. OEB acknowledges support from the Villum Foundation.

\section*{References}

%%%%%%%%%%%%%%%%%%%%%%%%%%%%%%%%%%%%%%%%%%%%%%%%%%%%%%%%%

\end{document}